\documentclass{raa}

\usepackage{graphicx}	
\usepackage{natbib}
\usepackage{amsmath}	
\usepackage{amssymb}	
\bibpunct{(}{)}{;}{a}{}{,}

\newcommand{\teff}{${T}_{\rm eff}$}
\newcommand{\logg}{$\log{g}$}

\usepackage[pagebackref=true]{hyperref}

\begin{document}
	
	\title{The Stellar Spectra Factory (SSF) Based On SLAM}
	
	\volnopage{Vol.0 (20xx) No.0, 000--000}      
	\setcounter{page}{1}          
	
	\author{Wei Ji 
		\inst{1,2}
		\and Chao Liu
		\inst{1,2,3}
		\and Bo Zhang
		\inst{1}
	}
	
	\institute{Key Laboratory of Space Astronomy and Technology, National Astronomical Observatories, CAS,  \\
		Beijing 100101, China; {\it jiwei@nao.cas.cn}\\
		\and
		University of Chinese Academy of Sciences,  \\
		Beijing 100049, China\\
		\and
		Institute for Frontier in Astronomy and Astrophysics, Beijing Normal University, Beijing 100875, China\\
		\vs\no
		{\small Received 20xx month day; accepted 20xx month day}}

	\abstract{An empirical stellar spectral library with large coverage of stellar parameters is essential for stellar population synthesis and studies of stellar evolution. In this work, we present Stellar Spectra Factory (SSF), a tool to generate empirical-based stellar spectra from arbitrary stellar atmospheric parameters. The relative flux-calibrated empirical spectra can be predicted by SSF given arbitrary effective temperature, surface gravity, and metallicity. SSF constructs the interpolation approach based on the Stellar LAbel Machine (SLAM), using ATLAS-A library, which contains spectra covering from O type to M type, as the training dataset. 
		SSF is composed of 4 data-driven sub-models to predict empirical stellar spectra. Sub-model SSF-N can generate spectra from A to K type and some M giant stars, covering $3700 < $\teff$ < 8700$ K,  $0 < $\logg$ < 6$ dex, and $ -1.5 < $[M/H]$ < 0.5$ dex. Sub-model SSF-gM is mainly used to predict M giant spectra with $3520 < $\teff$ < 4000$\,K and $ -1.5 < $[M/H]$ < 0.4$ dex. Sub-model SSF-dM is for generating M dwarf spectra with $3295 < $\teff$ < 4040$\,K, $ -1.0 < $[M/H]$ < 0.1$ dex. And sub-model SSF-B can predict B-type spectra with $9000 < $\teff$ < 24000$\,K and $ -5.2< M_G < 1.5 $ mag. The accuracy of the predicted spectra is validated by comparing the flux of predicted spectra to those with same stellar parameters selected from the known spectral libraries, MILES and MaStar. The averaged difference of flux over optical wavelength between the predicted spectra and the corresponding ones in MILES and MaStar is less than 5\%. More verification is conducted between the magnitudes calculated from the integration of the predicted spectra and the observations in PS1 and APASS bands with the same stellar parameters. No significant systematic difference is found between the predicted spectra and the photomatric observations. The uncertainty is 0.08\,mag in $r$ band for SSF-gM when comparing with the stars with the same stellar parameters selected from PS1. And the uncertainty becomes 0.31\,mag in $i$ band for SSF-dM when comparing with the stars with the same stellar parameters selected from APASS.
	}
	
	\authorrunning{W. Ji, C. Liu \& B. Zhang}            
	\titlerunning{The Stellar Spectra Factory (SSF) Based On SLAM}  
	
	\maketitle
	
	

	\section{Introduction}
	An empirical spectral library with complete stellar parameter coverage and flux calibration is of great significance for stellar population synthesis \citep{Bru2003,mar11,vaz10,vaz12}, which can be used to model stellar populations of galaxies \citep{2007IAUS..241...68P,2004A&A...425..881L}. In addition, an empirical spectral library can provide reference for classifying stars and determining stellar atmospheric parameters \citep{PRU01,kol09}. An automatic stellar parametrization and spectral classification tools, now a days, are necessary for large spectroscopic surveys, such as LAMOST \citep{2012RAA....12..735D}, SDSS/APOGEE \citep{Majewski2012}, GALAH \citep{2012ASPC..458..393F} etc. 
	
	Compared with theoretical spectral library, the spectra in an empirical library can better reflect the real information of stars so that it can also be used to constrain the theoretical stellar model. 
	The empirical stellar spectral libraries are widely used, such as Pickles \citep{pickl85,pickl98},ELODIE \citep{PRU01}, STELIB \citep{LEB03}, UVES-POP \citep{bagn03}, INDO-US \citep{vald04}, MILES \citep{sanch06} etc. The main shortcomings for these libraries is the poor coverage of parameters or the limited wavelength range by the capacity of observation instruments. 
	
	In recent years, newly released experiential spectral libraries, e.g. \cite{Du19}, MaStar \citep{yan19}, and ATLAS \citep{2023ApJS..265...61J}, are based on large spectroscopic surveys and hence cover broader parameter space and more stellar types. Compared with \cite{Du19} and MaStar, ATLAS covers larger parameter space and spectral types. In addition, the spectra of ATLAS calibrate in absolute manner, while spectra of MaStar calibrate the absolute fluxes that are normalized to 5450\AA.
	
	Other than the coverage of parameter space, all current stellar libraries provide discrete stellar spectra with more or less randomly distributed stellar parameters (as a comparison, the synthetic libraries usually align their stellar parameters with a regular parameter grid). To establish a synthetic stellar population, one always needs to develop an approach to interpolate spectra from the grid of data. As an empirical spectral library, a good interpolation may play important role in the population synthesis or in the stellar parametrization. 
	
	On the other hand, data-driven approaches for stellar parametrization from stellar spectra was well developed in these years. The Cannon~\citep{2015ApJ...808...16N}, SLAM~\citep{2020ApJS..246....9Z} etc. provide a forward modeling framework based on empirical stellar spectra. Such technique can take the place of the conventional interpolation to provide an alternative but effective way to obtain a spectrum with arbitrarily given stellar parameters. In this work, we apply the technique of forward-modeling developed by SLAM to provide a stellar spectra factory (SSF), that is, to produce an empirical stellar spectrum given an arbitrary set of stellar parameters as input. The SSF can be used as a software package for many potential usage, such as stellar population synthesis, stellar classification, and parametrization.
	
	
	This paper is organized as follows.  We give a brief description of SLAM in Section \ref{sec:method}, and we briefly describe the 4 training sets for SLAM to derive the four sub-models of SSF in Section \ref{sec:trainning set}. In Section \ref{sec:validation}, we assess the performance of the 4 sub-models in details. We list some caveats when using SSF in Section \ref{sec:discussion}. Finally, we draw the conclusions of this paper in Section \ref{sec:conclusion}.
	
	%
	%
	
	\section{ Method} \label{sec:method}
	
	The key of SSF is to set up a \emph{function} that maps a set of arbitrary stellar parameters to a stellar spectrum with somehow calibrated flux. SLAM is adopted as the function. SLAM is originally designed to estimate atmospheric parameters from the stellar spectra~\citep{2021ApJS..253...45L, 2021arXiv211006246G}. As a data-driven method, SLAM~\citep{2020ApJS..246....9Z} well combines the empirical stellar spectra with a wide range of atmospheric parameters and produce a forward model to generate spectrum in response to a set of given stellar parameters. Because the mapping from stellar parameters to the spectrum is highly non-linear, SLAM selects a machine-learning algorithm to handle it. Support Vector Regression \citep[SVR;][]{ss2004,Chang2011} is adopted as the default algorithm to build the mapping. It is a robust non-linear regression method used in many astronomical research and analysis \citep{ 2012MNRAS.426.2463L, 2014ApJ...790..110L, 2014ApJ...790..105L, 2015MNRAS.447..256B, 2015MNRAS.452.1394L}.
	%

	Support Vector Machine (SVM) is a kind of non-linear classifier that classifies data in a binary way according to a supervised learning process. As an extension of SVM, Support Vector Regression (SVR) slightly modifies SVM so that it can solve regression problems.

	As SLAM is based on SVR and adopts the radial basis function (RBF) as the kernel in SVR, it has three hyperparameters. $C$ and $\epsilon$ represent penalty level and tube radius, respectively, while $\gamma$ represents the width of the kernel. These three hyperparameters are all needed to be determined using the empirical spectra as the training data.
	
	We apply SLAM to build SSF in three steps:
	\begin{enumerate}
		
		\item [1.]\textbf{Pre-processing}. As the first step, we pre-process the training sample so that its spectral fluxes and parameters are in the standardized space. Unlike the other SLAM applications normalize the training spectra to remove the continua, we normalize the spectra to the sum of fluxes to keep the information of the relative flux calibration. Then, we standardize the training spectral fluxes at each wavelength pixel. The stellar labels, i.e. the stellar atmospheric parameters, are also standardized so that their mean is 0 and standard deviation is 1\footnote{The standardization is calculated as $x_{std}=(x-\bar{x})/\sigma_x$, where $\bar{x}$ and $\sigma_x$ are the mean and standard deviation, respectively, of the random variable $x$.}. The standardized procedure is taken from SLAM's internal algorithm. 
		
		

		\item [2.]\textbf{Data training}. We then train the SVR model with the standardized labels and fluxes at each wavelength pixel to obtain the best-fit hyperparameters for each wavelength pixel. 
		To achieve this, we firstly set the grid of hyperparameters to be $\epsilon$ = 0.05, $C$ = [0.01, 100] and $\gamma$ = [0.01, 100] and then run SLAM to search for the hyperparameters based on a $k$-fold cross-validation procedure, i.e., we separate the training data into $k$ groups, then we train the SVR model using $k-1$ groups and predict the spectra for the rest group. For each test, we measure the performance of the model with the mean squared error (MSE) at the $j$th pixel, which is defined as 
		\begin{equation} \label{MSE}
			MSE_{j} = \frac{1}{m} \sum_{i=1}^{m}[f_j(\vec{\theta_{i}}|\vec{\phi}) - f_{i,j}] ^ 2, 
		\end{equation}
		where $\vec{\theta_i}$ is the input stellar label vector of the $i$th star of the test data. $f_j(\vec{\theta_i}|\vec{\phi})$ denotes the predicted $j$th pixel of the test spectrum corresponding to $\vec{\theta_i}$ given the hyperparameter of $\vec{\phi}=(\epsilon, C, \gamma)$. And $f_{i,j}$ denotes the flux of the $j$th pixel of the $i$th spectrum of the test data. After $k$ run by selecting different group as the test data, the averaged $MSE_{j}$ of the $k$-fold cross-validation can be used to find the best-fit hyperparameters. We finally adopt the hyperparameters that can minimize the averaged $MSE_j$ for the $j$th wavelength pixel.
		
		\item[3.]\textbf{Spectra prediction}. The last step is to predict the spectrum corresponding to a set of given stellar labels with the trained SVR model with best-fit hyperparameters.
	\end{enumerate}

	\section{The training data set}
	\label{sec:trainning set}
	
	
	To provide a reliable and precise SSF, one needs a suitable empirical stellar spectral library to be acted as the training dataset of the SLAM. Some broadly used empirical library, e.g. MILES, ELODIE, BC03 and etc \citep{sanch06, PRU01, Bru2003} can only provide limited stellar spectra. Therefore, we select ATLAS-A library, which is a set of the ATLAS library \citep{2023ApJS..265...61J} and covers a wide range of spectral types, as the training data. ATLAS is an empirical stellar spectra library with resolution of $R\sim1800$ and wavelength coverage from 3800 to 8700 \AA.
	
	ATLAS-A contains 5342 spectra with effective temperature (\teff), surface gravity ($\log{g}$), and metallicity ([M/H]) and 242 spectra with only the effective temperature and surface gravity. It covers spectral types from O to M and also includes some special types of stars, such as A supergiant, blue horizontal-branch (BHB), and Carbon stars. The parameter coverage of ATLAS-A is from \teff$\sim3000$ to 50000 K, from \logg$\sim$0 to 6 dex, and from [M/H]$\sim$-1.5 to 0.5 dex. ATLAS-A also provides the coverage of absolute magnitude of $G$-band from ~ -9.5 to 12 mag. The spectra in ATLAS-A are absolutely calibrated in flux with wavelength range of 3800-8700\AA. 
	Because not all spectra in ATLAS-A have all three atmospheric parameters measured, we separate them into four groups and set up for SLAM sub-models.
	
	\subsection{Training data for SSF-N}
	We firstly select spectra with \teff, $\log{g}$ and [M/H] from ATLAS-A as the training set for SSF-N, i.e. the SSF for normal stars.
	The ranges of parameters of SSF-N are $3700<$\teff$<8700$ K, $0<$\logg$<6$ dex, and $-1.5<$[M/H]$<+0.5$ dex. 
	To ensure that the parameters are roughly evenly distributed, we assign the samples into small grids in the \teff, \logg, and [M/H] space with size of $\Delta T_{\rm eff}=100$ K, $\Delta \log g=0.2$ dex, and $\Delta{\rm[M/H]}=0.05$\,dex. In each grid, we select 1 to 3 spectra with highest signal-to-noise ratio in $g$-band. We finally select 3609 spectra as the training spectra, covering from A to early M-type stars. 
	
	\subsection{Training data for SSF-gM}
	We select M giant spectra only with \teff\ and [M/H] as the training data for SSF-gM. The [M/H] of M giants is measured by SLAM \citep{2023arXiv230308344Q}. This training set contains 249 spectra. 
	The range of \teff\ is from 3520 to 4000 K and that of [M/H] is from -1.5 to 0.4 dex.
	
	\subsection{Training data for SSF-dM}
	SSF-dM is trained with M dwarf spectra with only \teff\ and [M/H]. The [M/H] for M dwarfs is adopted from \cite{2021ApJS..253...45L}. This training set contains 217 spectra. The range of \teff\ is from 3295 to 4040 K and that of [M/H] is from -1.0 to 0.1 dex.
	
	\subsection{Training data for SSF-B}
	Because some parameters, such as metallicity, of B-type stars are not given, we use \teff\ and absolute magnitude ($M_G$) as input parameters and 30 B type spectra as input spectra for SSF-B. The range of \teff\ is from 9000 to 24000 K and $M_G$ is from -5.2 to 1.5 mag.
	%
	%

	\section{Results}
	\label{sec:validation}
	
	According to the different training sets, we finally establish the SSF, which is composed of four sub-models for different spectral types: SSF-N, SSF-gM, SSF-dM and SSF-B. The coverage of spectral types and parameters for the 4 sub-models are listed in Table \ref{table:names}. All four sub-models can be downloaded from the website: \url{https://nadc.china-vo.org/res/r101182/} and we also provide a \emph{README} which can be downloaded from the website: \url{https://github.com/hypergravity/SFF} to introduce how to use the models. The source code of SLAM can be downloaded on GitHub \url{https:// github.com/hypergravity/astroslam} under an MIT License.
	
	\begin{table*}
		\caption{The coverage of spectral type and parameters of the SSF packages.}
		\centering
		\begin{tabular}{c|c|c}
			\hline
			Library name & Coverage of spectral type & Coverage of parameters                                                                                           \\ \hline
			SSF-N    & A, F, G, K, and early-M type (part of M giant)                  & \begin{tabular}[c]{@{}c@{}}Teff:  3700 to 8700 K;\\ logg: 0 to 6 dex;\\ {[}M/H{]}: -1.5 to 0.5 dex.\end{tabular} \\ \hline
			SSF-gM    & M giant                   & \begin{tabular}[c]{@{}c@{}}Teff:  3520 to 4000 K;\\ {[}M/H{]}: -1.5 to 0.4 dex;\\ giant: 1\end{tabular}          \\ \hline
			SSF-dM    & M dwarf                   & \begin{tabular}[c]{@{}c@{}}Teff:  3295 to 4040 K;\\ {[}M/H{]}: -1.0 to 0.1 dex;\\ dwarf: 0\end{tabular}          \\ \hline
			SSF-B    & B type                    & \begin{tabular}[c]{@{}c@{}}Teff:  9000 to 24000 K;\\ $M_G$: -5.2 to 1.5 dex\end{tabular}                          \\ \hline
		\end{tabular}
		\label{table:names}
	\end{table*}
	
	\subsection{Validation}
	In order to assess the accuracy of the predicted spectra, we verify them from two aspects. The first is to verify the flux continua of predicted spectra by comparing with the spectra corresponding to the same parameters in the known stellar empirical spectra library. The second is to verify the accuracy of the photometry calculated from the predicted spectra by comparing with the observed magnitude with same stellar parameters. It is noted that for different training sample sets, we adopt different verification methods.
	
	\subsubsection{Validation of the spectra}
	\label{sect: validate_library}
	
	In order to verify the predicted spectra, we make a few comparisons with spectra. Firstly, we select the normal spectral type spectra with stellar labels from the known libraries. Then, we input the stellar labels of these selected spectra to SSF to derive the corresponding predicted spectra.
	
	We select the spectra from MILES and MaStar libraries to verify SSF-N and SSF-gM, while we only select spectra from MILES library for the verification of SSF-dM, since MILES contains few M-type stars. 
	
	
	First, we predict the spectra from SSF using the stellar labels of the spectra in the known libraries. Then, since MaStar does not correct the extinction, we need to add extinction effect back to the predicted spectra before comparison. This step can be conducted as
	\begin{equation}\label{eq:extiction}
		F_{\rm ex}(\lambda)= F_{\rm pre}(\lambda)10^{-A_VR(\lambda)},
	\end{equation}
	where $F_{ex}(\lambda)$ represents the flux of predicted spectra with extinction effect added back, $F_{pre}(\lambda)$ denotes the flux of predicted spectra, ${A_V}$ is $V$ band extinction obtained from SED fitting of the photometries of the selected stars from MaStar and $R(\lambda)$ is the extinction law adopted from \citet{car89}.
	
	
	Finally, we normalize both the predicted spectra and the spectra from the known libraries using the sum of flux of the spectra over all range of wavelength. Since that the spectra in MILES are originally normalized to the median flux in the range of 5000 to 5050\,{\AA} while spectra with MaStar are originally normalized to the flux at 5450\,{\AA}, we perform a further normalization of the predicted spectra before comparison. When comparing the predicted spectra with MILES spectra, we further normalize both spectra by the median flux in the range of 5000 to 5050\,{\AA}. When comparing the predicted spectra with MaStar spectra, we further normalize both spectra using the flux at 5450\,{\AA}. We use two steps of normalization to adopt a more accurate flux comparison for that the accuracy of comparison with MaStar can be improved by 19\% if we add one more normalization at 5450\,{\AA} after the first normalization.  
	
	In order to show the comparison more clearly, we quantified the comparison results by calculating the flux ratio over the wavelength. At each wavelength pixel, we calculate the 15th, 50th, and 85th percentiles. The difference between 15th and 50th percentiles and between 50th and 85th percentiles in each wavelength pixel are calculated by:
	%
	\begin{align}\label{eq: COMPARISON RESIDUAL}
		&\Delta F_{15-50}(\lambda) =\frac {F_{15}(\lambda) - F_{50}(\lambda)} {F_{50}(\lambda)}\nonumber\\
		&\Delta F_{50-85}(\lambda) = \frac{F_{50} (\lambda)- F_{85}(\lambda)} {F_{50}(\lambda)},
	\end{align}
	where $\Delta F_{15-50}(\lambda)$ and $\Delta F_{50-85}(\lambda)$ denotes the difference of flux ratio between 15th and 50th percentiles and between 50th and 85th percentiles, respectively. $F_{15}(\lambda)$, $F_{50}(\lambda)$ and $F_{85}(\lambda)$ represents the value of the flux ratio of multiple spectra at 15th, 50th, and 85th percentiles, respectively, at each wavelength pixel. Notice that the fluxes of strong absorption lines are excluded in the calculation.
	
\begin{figure*}
		\begin{center}
			\includegraphics[width=1.0\textwidth]{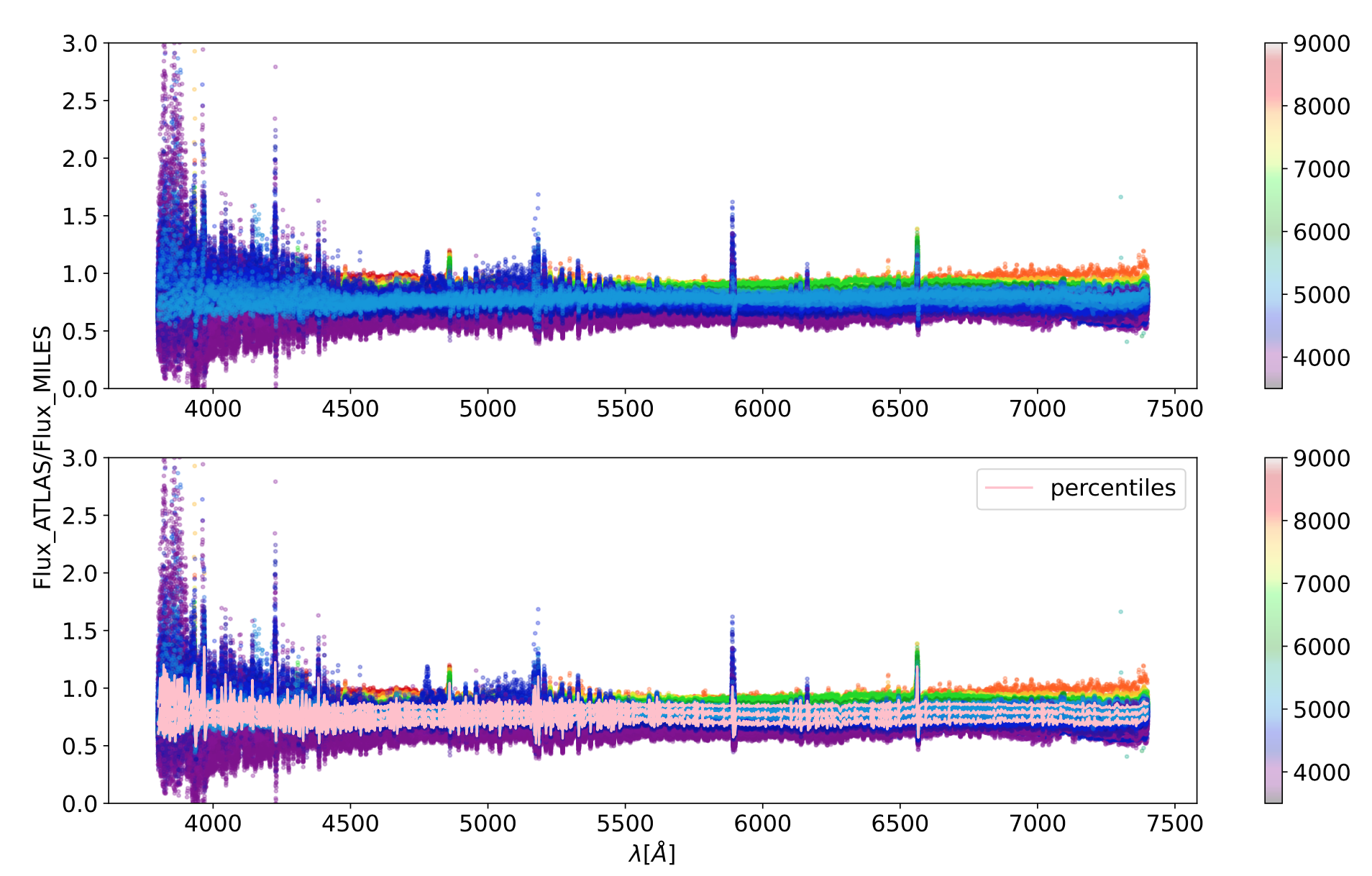}
			\caption{Flux ratio between predicted spectra from SSF-N and the corresponding MILES spectra with the same stellar parameters. In the upper panel, colors represent the effective temperatures of the individual spectrum. In the bottom panel, the pink thick lines indicate the 15th, 50th, and 85th percentiles of the flux ratio at each wavelength pixel.}
			\label{fig:miles_comparison}
		\end{center}
	\end{figure*}

	We use 532 spectra from MILES library and 237 spectra from MaStar library for the comparison with SSF-N. And we use 5 and 4 spectra selected from MILES library for the comparison with SSF-gM and SSF-dM, respectively.
	Figure~ \ref{fig:miles_comparison} shows the distribution of the flux ratio between the predicted spectra from SSF-N and the corresponding selected MILES spectra with same stellar parameters. The color codes the effective temperatures of individual stars in the upper panel. The pink lines in the bottom panel denote the 15th, 50th, and 85th percentiles of the flux ratios. The mean values over wavelength of $\Delta F_{15-50}(\lambda)$ and $\Delta F_{50-85}(\lambda)$ are 3.7\% and 4.2\%, respectively.
	
\begin{figure*}
		\begin{center}
			\includegraphics[width=1.0\textwidth]{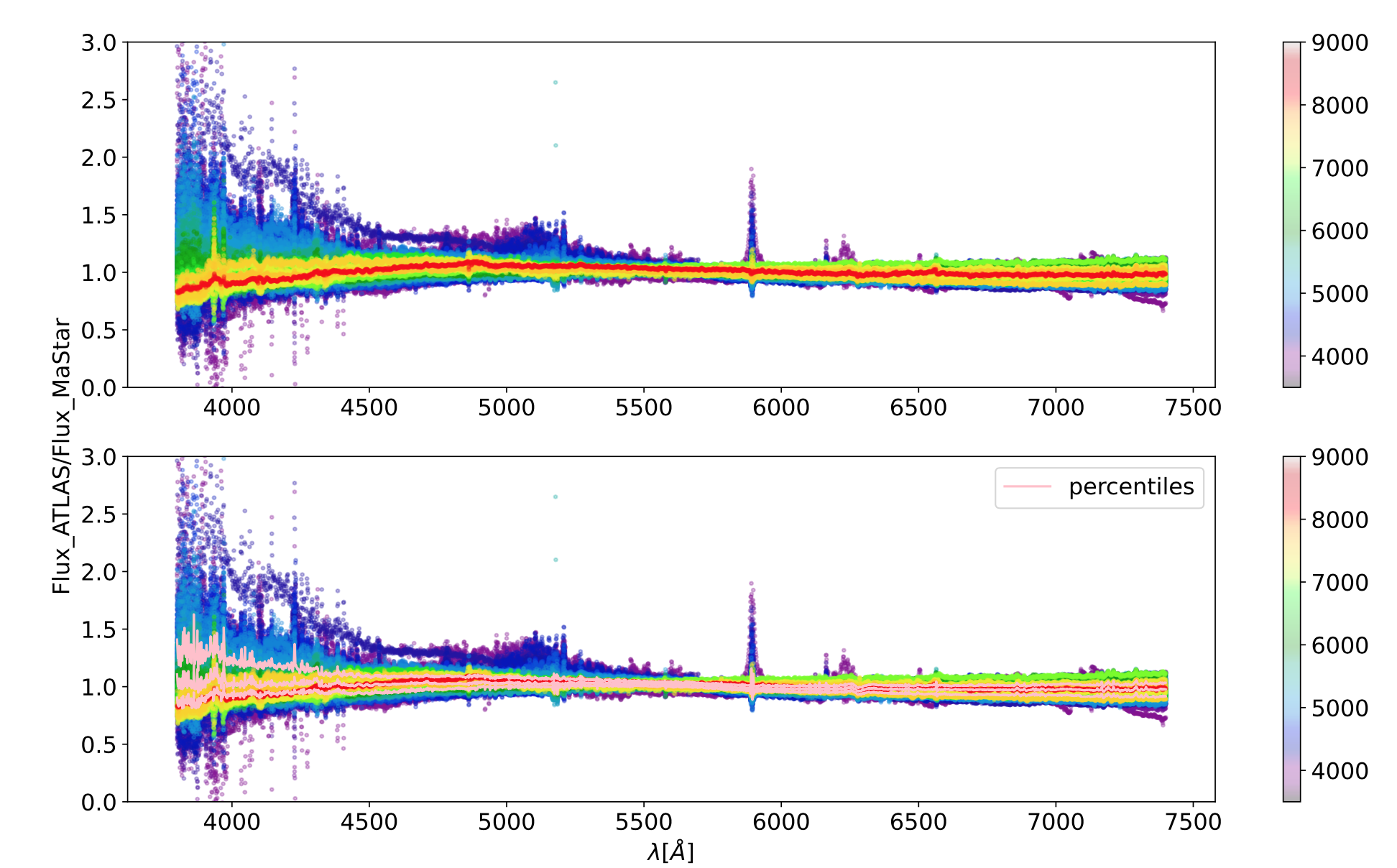}
			\caption{Flux ratio between predicted spectra from SSF-N and the corresponding MaStar spectra. The symbols and colors are same as in Figure~\ref{fig:miles_comparison}.}
			\label{fig:mastar_comparison}
		\end{center}
	\end{figure*}

	Figure~\ref{fig:mastar_comparison} shows the distribution of the flux ratio between the predicted spectra from SSF-N and corresponding spectra with same stellar parameters selected from MaStar. The mean values over the wavelength of the flux ratios $\Delta F_{15-50}(\lambda)$ and $\Delta F_{50-85}(\lambda)$ are 4.4\% and 5.0\%.
	
\begin{figure*}
		\begin{center}
			\includegraphics[width=1.0\textwidth]{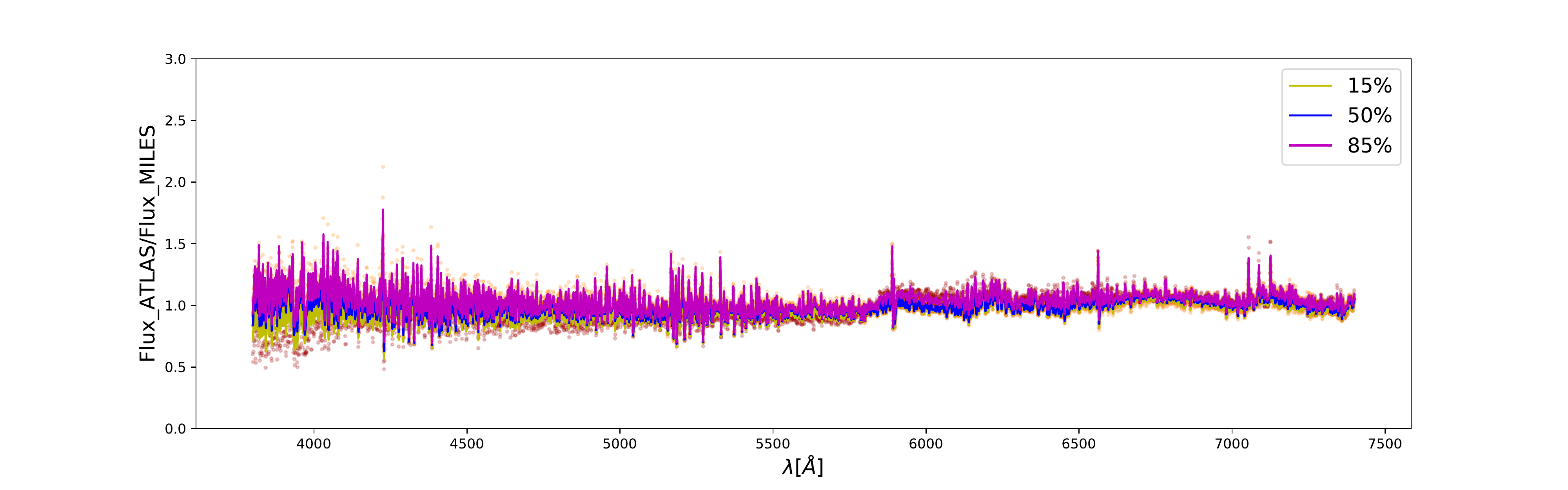}
			\caption{Flux ratio between predicted spectra from SSF-gM and the corresponding MILES spectra. The yellow, blue, and purple thick lines indicate the 15th, 50th, and 85th percentiles of the flux ratio at each wavelength pixel, respectively.}
			\label{fig:mg_miles_comparison}
		\end{center}
	\end{figure*}

\begin{figure*}
		\begin{center}
			\includegraphics[width=1.0\textwidth]{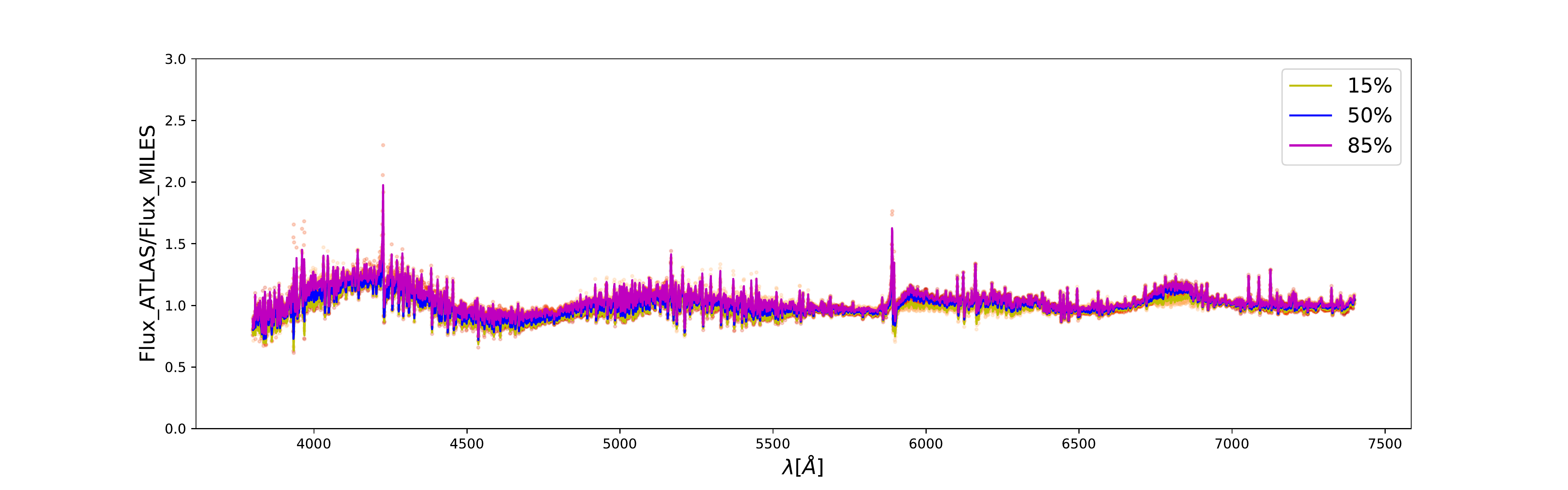}
			\caption{Flux ratio between predicted spectra from SSF-dM and the corresponding MILES spectra. The symbols and colors are same as in Figure~\ref{fig:mg_miles_comparison}.}
			\label{fig:md_miles_comparison}
		\end{center}
	\end{figure*}

	Since either SSF-gM or SSF-dM contains only one spectral type of stars (M giant and M dwarf spectra, respectively), we selected 5 M giants and 4 M dwarf stars in MILES library for the comparison. Figures~\ref{fig:mg_miles_comparison} and~\ref{fig:md_miles_comparison} show the flux ratios between the predicted spectra by SSF-gM and SSF-dM and the corresponding MILES spectra. 
	It is noted that some MILES spectra of cooler M type stars is not consistent with their originally published stellar parameters. Therefore, we adopt the \teff and metallicity from \cite{2016A&A...585A..64S} as the atmospheric parameters to extract spectra from SSF-gM or SSF-dM. The mean values over wavelength of $\Delta F_{15-50}(\lambda)$ and $\Delta F_{50-85}(\lambda)$ are 3.0\% and 3.6\%, respectively, for SSF-gM, while the values become 2.7\% and 2.9\%, respectively, for SSF-dM. 
	
\begin{figure*}
		\begin{center}
			\includegraphics[width=1.0\textwidth]{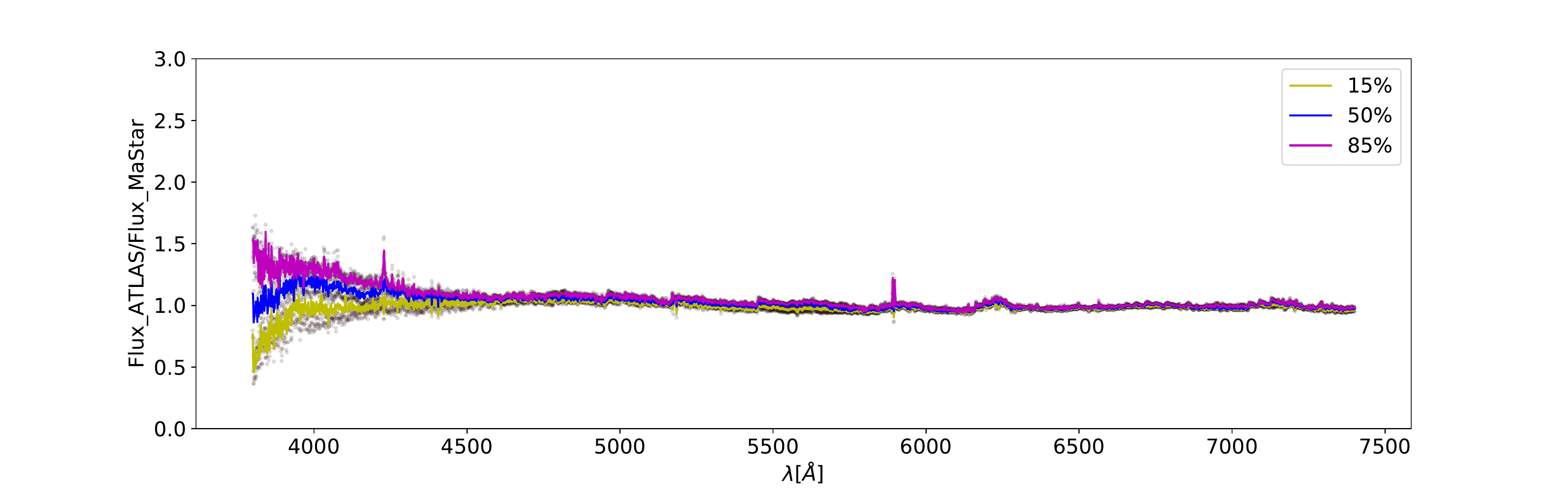}
			\caption{Flux ratio between predicted spectra from SSF-gM and the corresponding MaStar spectra. The symbols and colors are same as in Figure~\ref{fig:mg_miles_comparison}.}
			\label{fig:mg_mastar_comparison}
		\end{center}
	\end{figure*}

	
	We also find 4 common M giants in both MaStar library and ATLAS-A. We obtain the predicted spectra from SSF-gM for these stars using the stellar parameters given by MaStars. The flux ratios of these spectra are shown in Figure~\ref{fig:mg_mastar_comparison}. The mean values of $\Delta F_{15-50}(\lambda)$ and $\Delta F_{50-85}(\lambda)$ are 2.7\% and 2.3\%, respectively. The dispersion of the flux ratio becomes larger at wavelength range smaller than 4500\AA ~in Figure~\ref{fig:mg_mastar_comparison}.
	
\begin{figure*}
		\begin{center}
			\includegraphics[width=1\textwidth]{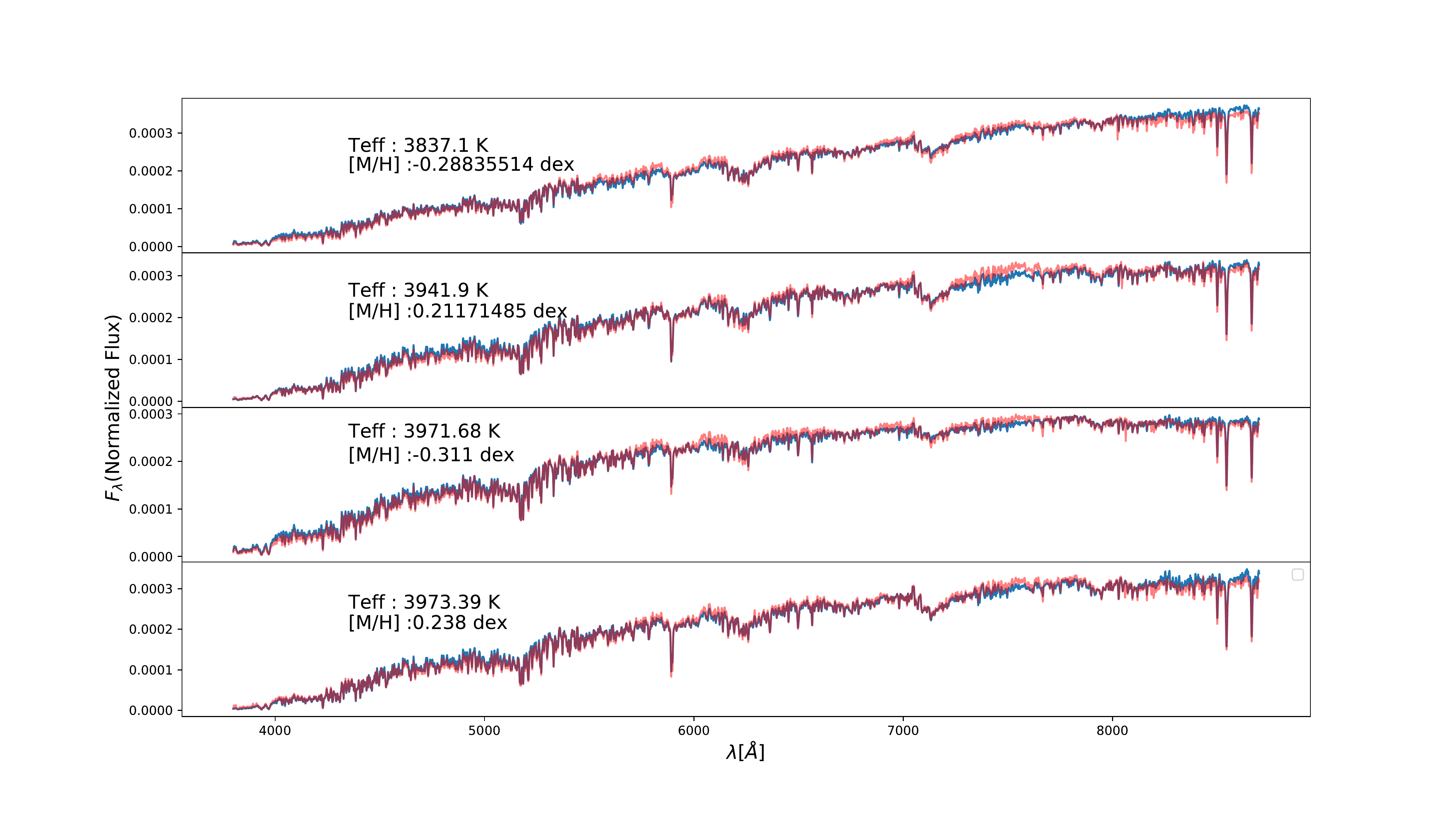}
			\caption{This figure shows some samples of the direct comparison between the MaStar spectra and the corresponding predicted spectra from SSF-gM. The blue and red indicate the predicted spectra from SSF and the spectra from MaStar, respectively. We label the different parameters in the panels for each spectra.}
			\label{fig:Mg_spectra_compare}
		\end{center}
	\end{figure*}

	We find the largest uncertainty occurs in the wavelength range of 3800-4000 \AA, which is illustrated in Figure~\ref {fig:mg_mastar_comparison}. To see these differences more clearly, we plot a few typical sample spectra in Figure~\ref {fig:Mg_spectra_compare}. The blue lines denote the predicted spectra and the red lines denote the spectra with same stellar parameters selected from MaStar spectra. All the spectra are normalized to the flux at 5450 \,{\AA}. At 3800-4000\,\AA, the large uncertainty shown in Figure ~\ref {fig:mg_mastar_comparison} is likely due to the lower fluxes at blue wavelength for these cold stars.
	
\begin{figure*}
		\begin{center}
			\includegraphics[width=1\textwidth]{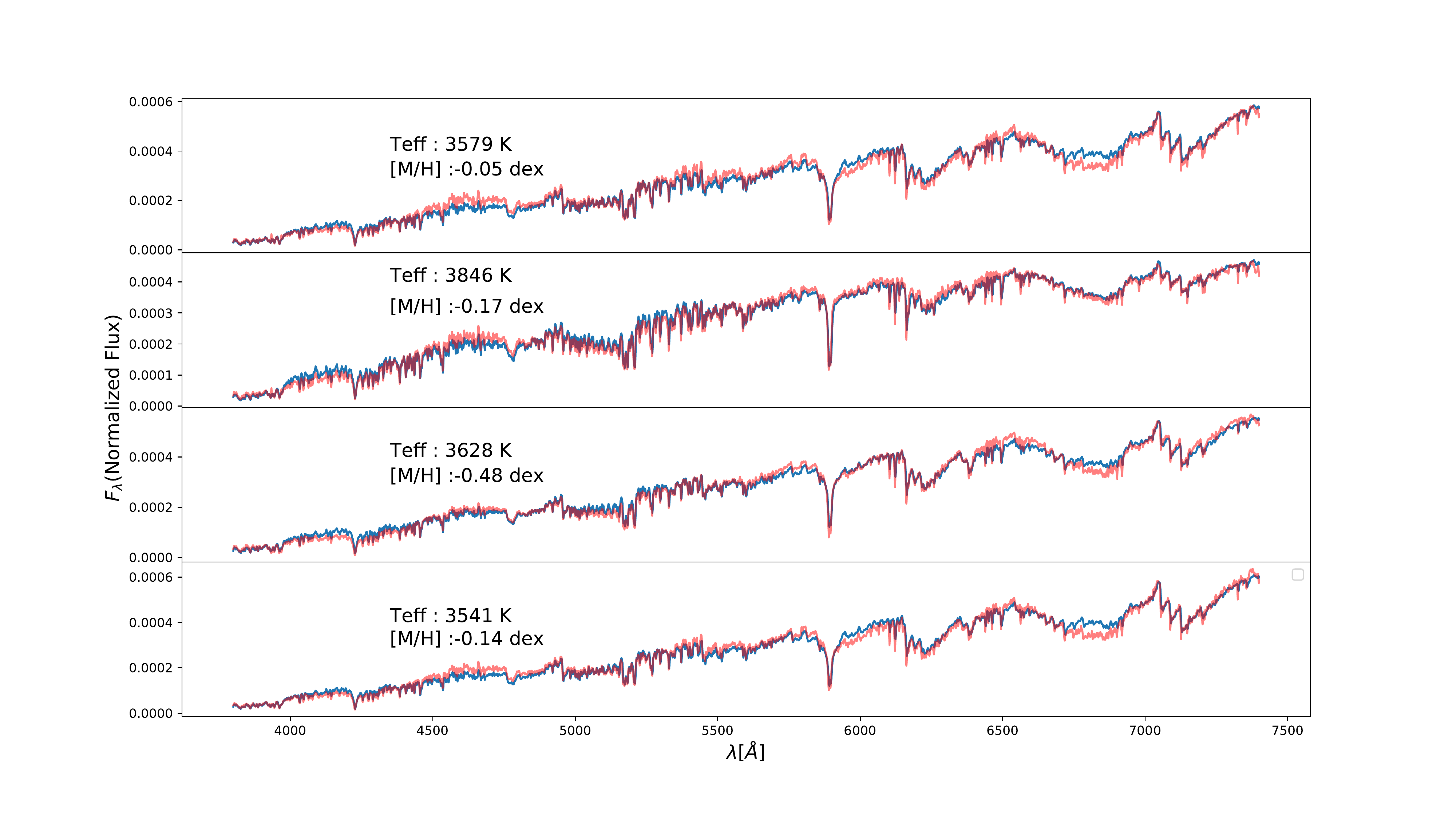}
			\caption{This figure shows some samples of the direct comparison between the MILES spectra and the corresponding predicted spectra from SSF-dM. The blue and red indicate the predicted spectra from SSF and the spectra from MILES, respectively. We label the different parameters in the panels for each spectra.}
			\label{fig:Md_miles_spectra_compare}
		\end{center}
	\end{figure*}

\begin{figure*}
		\begin{center}
			\includegraphics[width=1\textwidth]{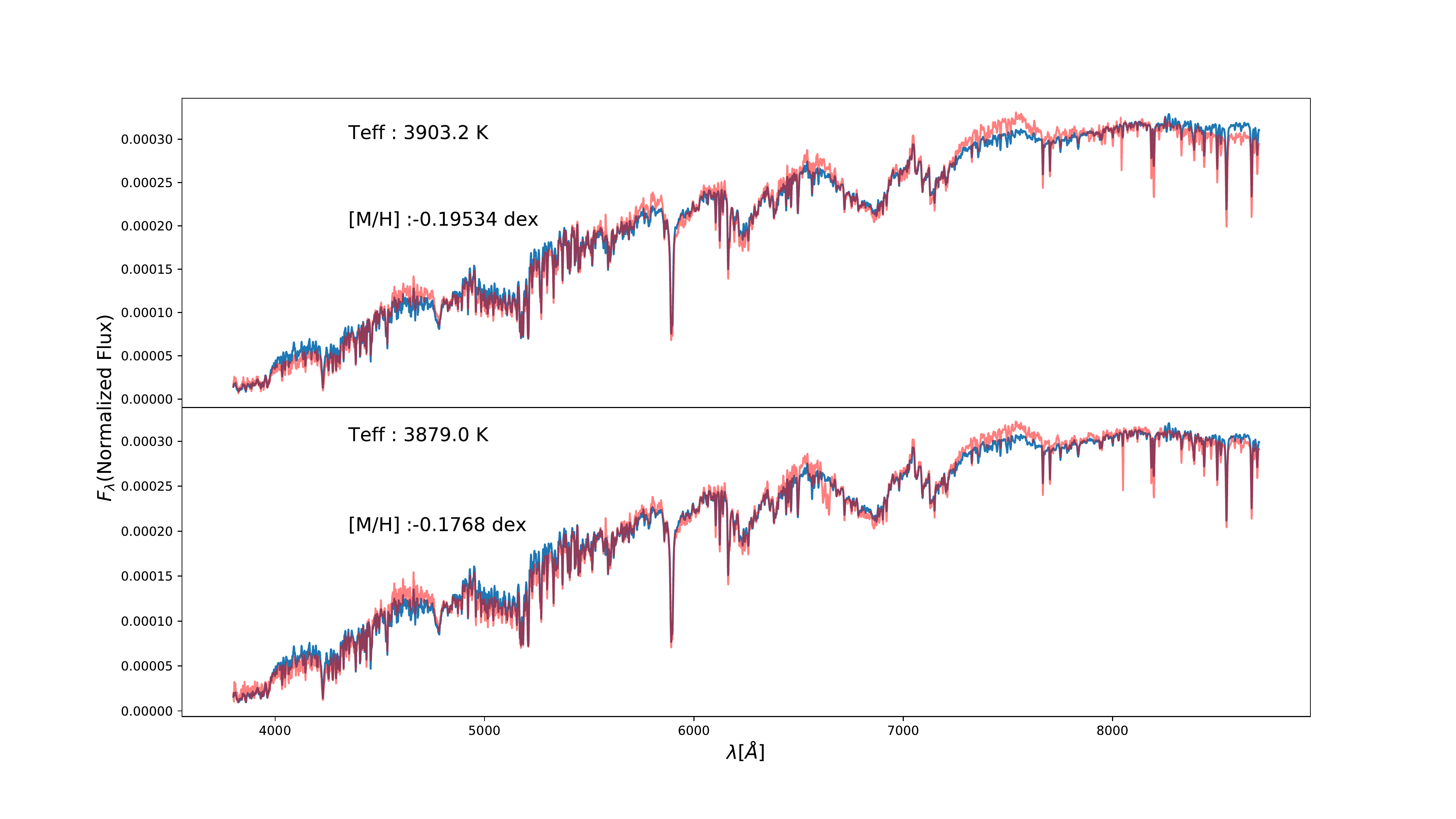}
			\caption{This figure shows some samples of the direct comparison between the MaStar spectra and the corresponding predicted spectra from SSF-dM. The blue and red indicate the predicted spectra from SSF and the spectra from MaStar, respectively. We label the different parameters in the panels for each spectra.}
			\label{fig:Md_mastar_spectra_compare}
		\end{center}
	\end{figure*}

	
	For SSF-dM, we find the largest uncertainty occur in a wide range of wavelength, as shown in the Figure~\ref {fig:md_miles_comparison}. Similarly, we plot a comparison some predicted sample spectra with the MILES spectra with same stellar parameters in Figure~\ref {fig:Md_miles_spectra_compare}. The spectra shown in the figure are normalized to the median flux in the range of  5000 to 5050 \AA. We find that the largest difference in wavelength range of 6750-6850 \AA~ mainly appear in the M dwarf spectra with \teff~less than 3700 K. It may be cause that either the absorption lines for atmosrpheric oxygen molecule in MILES is not cleanly reduced or they have been over-subtracted in the process of skylight subtraction of LAMOST data pipeline. To further investigate this difference, we compare some sample M dwarf spectra from MaStar spectral library with the predicted spectra using same stellar parameters. As shown in Figure~\ref {fig:Md_mastar_spectra_compare}, the predicted spectra are well consistent with the MaStar spectra. This means that it is likely the MILES that does not cleanly subtract the atmospheric O$_2$ lines during data reduction. Except the uncertainty at 6750-6850 \AA, Figures~\ref{fig:Md_miles_spectra_compare} and \ref{fig:Md_mastar_spectra_compare} show that the wave-like uncertainties are mainly due to  a) the larger difference of continuum between the predicted and MILES/MaStar spectra or b) the large uncertainties in stellar parametrs for these cold stars.
	
	

	
	\subsubsection{Validation by the observed photometric measurement}
	\label{sec:comparison by photometric}
	
	To further verify the accuracy of the predicted spectra, we also compare the values of the observed magnitude with the corresponding ones that calculated from the predicted spectra. In accordance with the steps in section~\ref{sect: validate_library}, we first select the samples with stellar labels falling in the parameter space of the training set from LAMOST DR5 data. Then, we apply the SVR model to find the corresponding predicted spectra using the stellar parameters provided by LAMOST catalog. We add extinction, which is derived from ATLAS-A catalog using SED fitting, back to the predicted spectra by Eq.~\ref{eq:extiction} and integrate the fluxes of the predicted spectra in each photometric band by 
	\begin{equation}\label{eq:fluxmag}
		F_{i} = \int{F_{\rm ex}(\lambda)T_i(\lambda)d\lambda},
	\end{equation}
	where $i$ denotes the name of band, $F_i$ is the derived flux of the given band from the predicted spectrum, and $T_i(\lambda)$ represents the transmission curve of the band $i$. The $F_i$ is then converted into AB magnitude by
	\begin{equation}\label{eq:magnitude calculate}
		m_{i} = -2.5\log \frac {F_{i}\times {c}}{{\lambda}^{2}}-48.6,
	\end{equation}
	where $m_{i}$ represents the AB magnitude, $\lambda$ is the effective wavelength for the given photometric band and $c$ is the speed of light.
	
	Because that the fluxes of ATLAS spectrum are calibrated according to $g$, $r$, and $i$ bands in PS1 \citep[The Panoramic Survey Telescope and Rapid Response System-1][]{cha16} and APASS \citep[The American Association of Variable Star Observers Photometric All-Sky Survey][]{2014CoSka..43..518H, 2016yCat.2336....0H}, we also select the photometric values of the same bands in these two photometric systems for comparison. 
	
	Since our predicted spectrum is not absolutely flux-calibrated, in order to obtain accurate magnitude value in photometric bands, we need to calculate the absolute flux of spectrum with extinction by
	\begin{equation}\label{eq:flux_k}
			F_{\rm cal}(\lambda) = k\times{F_{\rm ex}(\lambda)},
		\end{equation}
		where $F_{\rm cal}(\lambda)$ denotes the flux with extinction and distance that to calculate the observed magnitude and $k$ is coefficient between $F_{\rm cal}(\lambda)$ and $F_{\rm ex}(\lambda)$. 
		
		In order to get $k$, we firstly calculate the magnitude in $g$ band for $F_{\rm ex}(\lambda)$ and then we calculate the difference between the calculated magnitude and the observed magnitude in $g$ band. According to the Pogson formula, we get k by
		\begin{align}\label{eq:flux_k_cal}
			&\Delta g = {g_{\rm pre} - g_{\rm obs}}\\
			&k = 10^{-2.5\times{\Delta g}}\nonumber
		\end{align}
		where $g_{\rm pre}$ represents the magnitude for the predicted spectrum with extinction effect added and $g_{\rm obs}$ represents the observed magnitude in $g$ band.
		
		Since we use $g$ band to obtain $k$, the value of magnitude calculated from $F_{cal}(\lambda)$ in $g$ band must be equal to the observed magnitude. Therefore, we only choose $r$ and $i$ bands in PS1 and APASS photometric systems for comparison.
		
		For the validation of SSF-N, we generate two sets of samples with atmospheric parameters and spectra published by LAMOST DR5. One has been observed in PS1 with accurate (but not saturated) $g$, $r$, and $i$ bands and the other has APASS $g$, $r$, and $i$ bands. We randomly select 726 objects with clear photometry information and with representative observed parameter ranges in SSF-N. We finally obtain 484 samples with PS1 photometry and 242 samples with APASS photometry. Both samples are sent to SSF-N to obtain the predicted spectra by using their corresponding \teff, $\log{g}$, and $[M/H]$ from LAMOST as the input stellar labels. 
		
\begin{figure*}
			\begin{center}
				\includegraphics[width=1.\textwidth]{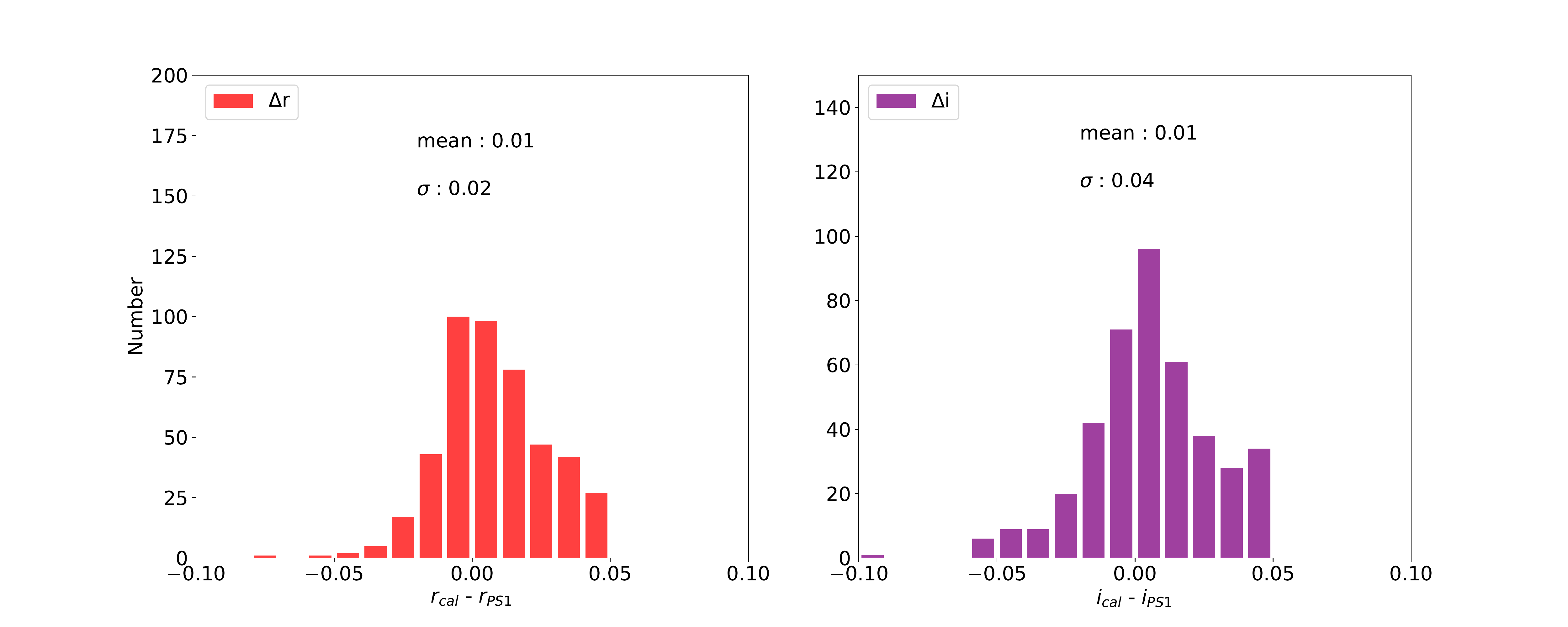}
				\caption{From left to right, the panels display the distributions of the difference between the observed and the SSF-N predicted magnitudes in PS1 $r$ and $i$ bands, respectively. The vertical axes are the star numbers.}
				\label{fig:gri_PS1_compare}
			\end{center}
		\end{figure*}

\begin{figure*}
			\begin{center}
				\includegraphics[width=1\textwidth]{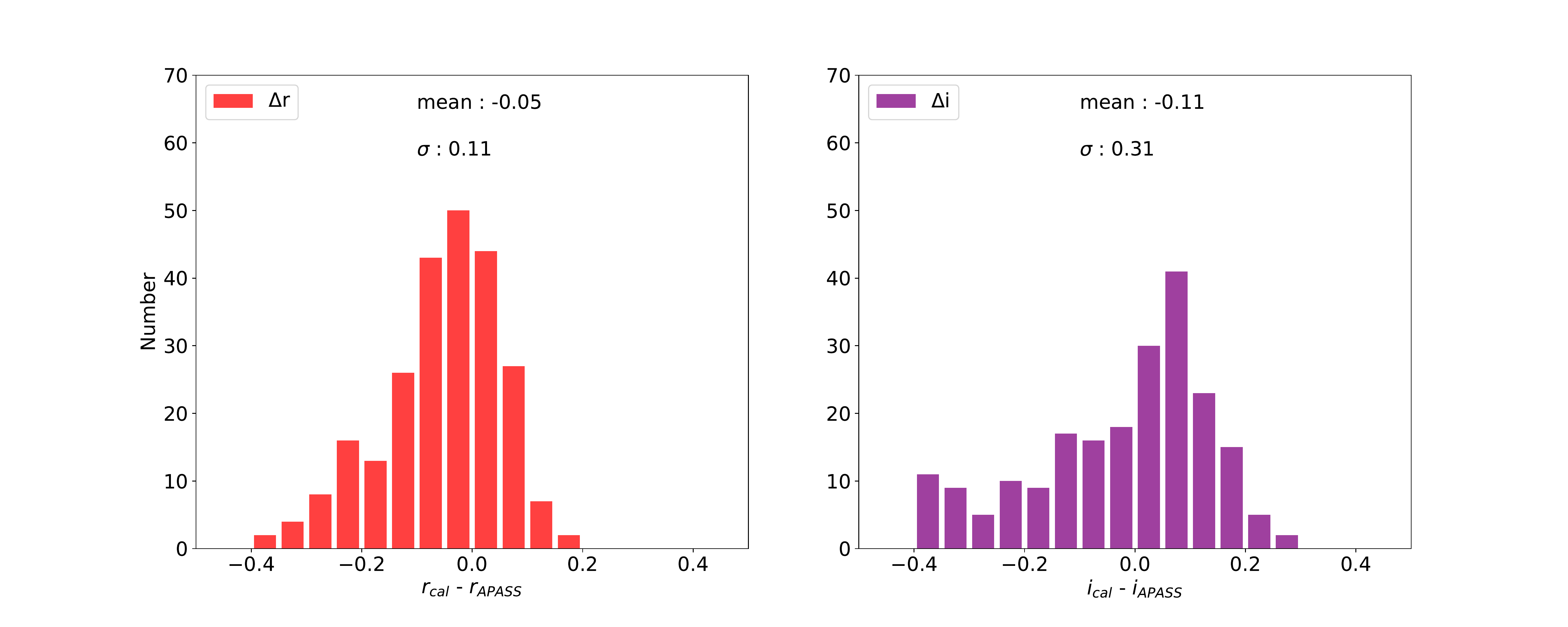}
				\caption{These panels are similar to Figure~\ref{fig:gri_PS1_compare}, but are compared with APASS  $r$ and $i$ bands.}
				\label{fig:gri_APASS_compare}
			\end{center}
		\end{figure*}

		Figures~\ref {fig:gri_PS1_compare} and~\ref{fig:gri_APASS_compare} display the distributions of the difference of magnitudes between the predicted spectral fluxes and observed magnitudes in $r$ and $i$ bands. In Figure~\ref{fig:gri_PS1_compare}, by comparing with the corresponding magnitudes from PS1, the systematic shifts of the predicted spectra are 0.01 and 0.01 mag in $r$ and $i$ band, respectively, with the random uncertainties of $<0.04$ mag, which is slightly larger than the typical error of PS1 photometry. In Figure~\ref{fig:gri_APASS_compare}, by comparing with APASS magnitudes, the systematic bias of the predicted spectra are -0.05 and -0.11 mag in $r$ and $i$ bands, respectively, with the random uncertainties less than 0.31 mag. Comparing with the typical error of APASS magnitude, which is around 0.1$\sim0.2$ mag, the uncertainty is not surprisingly large. The two figures show that the performance of the fluxes of spectra is poor in $i$ band, which is mainly contributed by the late type stars with significant molecular absorption bands in red wavelength. Nevertheless, these figures show that the prediction of spectra according to the stellar labels can robustly reproduce the photometry of stars with slightly larger uncertainty than observed photometry.
		
\begin{figure*}
			\begin{center}
				\includegraphics[width=1.\textwidth]{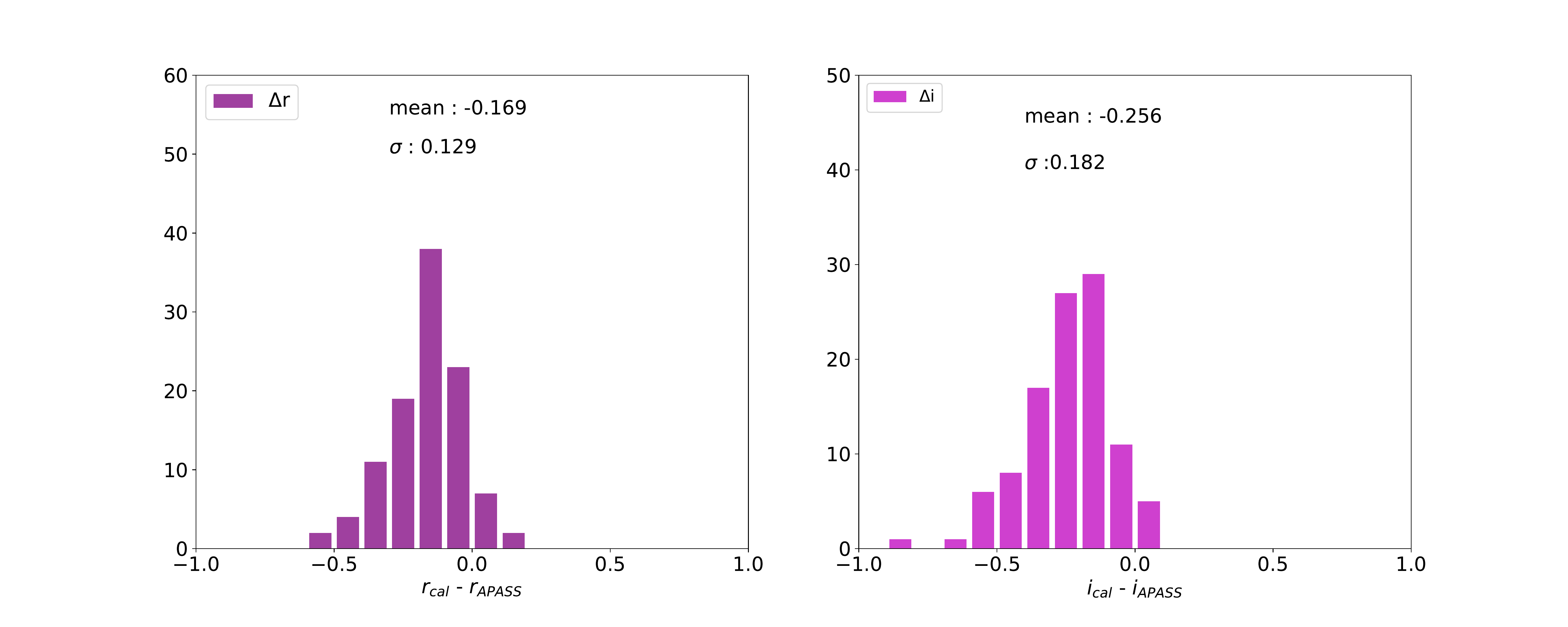}
				\caption{The histograms display the distributions of the difference between the observed and the SSF-gM predicted magnitudes in APASS $r$ (left panel) and $i$ (right panel) bands.}
				\label{fig:gri_mg_apass_compare}
			\end{center}
		\end{figure*}

\begin{figure*}
			\begin{center}
				\includegraphics[width=1\textwidth]{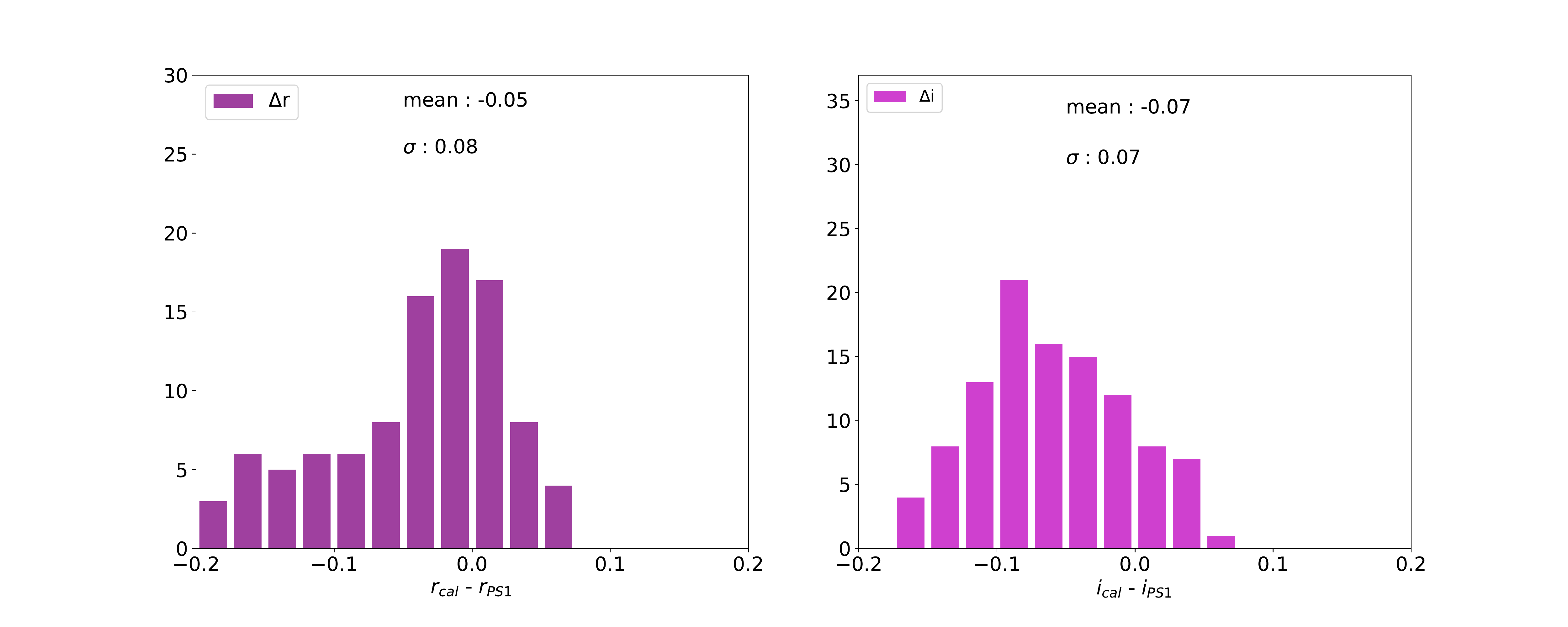}
				\caption{These panels are similar to Figure~\ref{fig:gri_PS1_compare}, but for SSF-dM. From left to right, the panels show the distributions of the difference between the observed and the predicted magnitudes in PS1 $r$ and $i$ bands, respectively.}
				\label{fig:md_gri_ps1_compare}
			\end{center}
		\end{figure*}

		Since the number of M-type training stars are relatively small, we select 106 samples with photometric information in APASS $g$, $r$, and $i$ bands from LAMOST DR5 data for the validation of SSF-gM and 108 samples with photometric information in PS1 $g$, $r$, and $i$ bands from LAMOST DR5 data for the validation of SSF-dM. 
		Figure~\ref {fig:gri_mg_apass_compare} displays the distribution of the difference between the observed and the SSF-gM predicted magnitudes in APASS $r$ and $i$ bands, while Figure~\ref{fig:md_gri_ps1_compare} displays the distribution of the difference between the observed and the SSF-dM predicted magnitudes in PS1 $r$ and $i$ bands. It is seen that the systematic bias are -0.169 and -0.256 mag in the comparison with APASS $r$ and $i$ band, respectively, with the random uncertainties of less than 0.18 mag in Figure~\ref {fig:gri_mg_apass_compare}. And the systematic shift are -0.05 and -0.07 mag in the comparison with PS1 $r$ and $i$ band, respectively, with the random uncertainties of less than 0.08 mag in Figure~\ref {fig:md_gri_ps1_compare}. The bias in APASS $i$ band shows in Figure~\ref {fig:gri_mg_apass_compare} is slightly large, which is probably due to the fact that the parameters of the training samples do not sufficiently fill in with the parameter space and the photometry of the red stars may not be precise in APASS.
		
\begin{figure*}
			\begin{center}
				\includegraphics[width=1\textwidth]{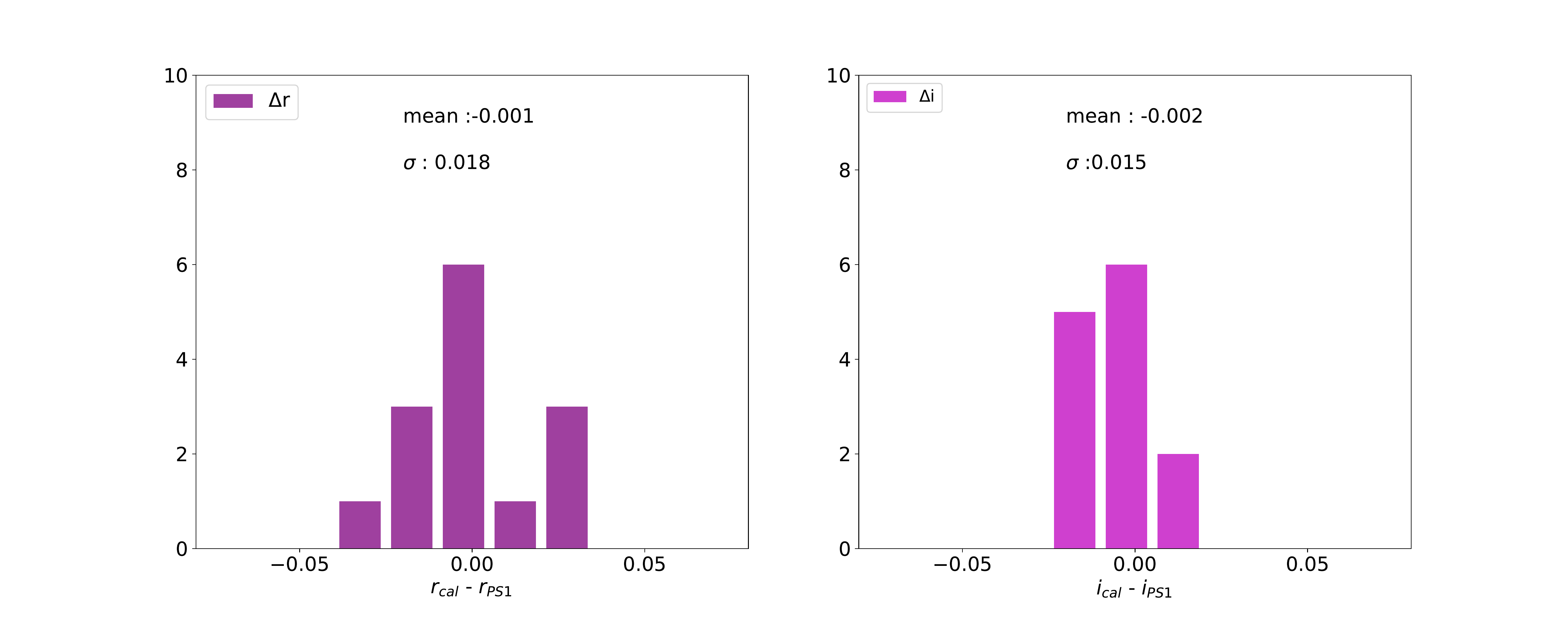}
				\caption{These panels are similar to Figure~\ref{fig:gri_PS1_compare}, but for SSF-B. From left to right, the panels show the distributions of the difference between the observed and the predicted magnitudes in PS1 $r$ and $i$ bands, respectively.}
				\label{fig:OB_gri_ps1_compare}
			\end{center}
		\end{figure*}

		
		We select 14 spectra with input parameters in the training samples of SSF-B to evaluate the difference between the observed magnitude and the corresponding magnitude from the spectra predicted by SSF-B. Figure~\ref{fig:OB_gri_ps1_compare} shows the final comparison result. The bias in the figure are -0.001 and -0.003 mag in PS1 $r$ and $i$ band, respectively with the uncertainties are 0.018 and 0.015 mag, respectively. 
		
		%
		

		\subsection{Caveats}\label{sec:discussion}
		There are a few caveats need to be noticed when using SSF. 
		\begin{enumerate}
			\item With the exception of SSF-N, which adopts three parameters (\teff, $\log{g}$ and [M/H]) as stellar labels, the rest of the libraries basically contain only two. However, to minimize the change of the SLAM code, we adopt a constant value 1 and 0 as the fixed third parameter for SSF-gM and SSF-dM, respectively. The SSF-N covers the spectral types from A to early-M,  while the M type stars here are only some of the early type M giants. And the rest 3 libraries are only contain one specific class of stellar spectra, so users need to pay attention before applying these libraries.
			\item According to the stellar parameters coverage in the training samples, the spectral prediction at the edge of the training sample may not be accurate. This is mainly cause that there are very few stars are located near the edge of the training sample.

		\end{enumerate}

		\section{Conclusions}
		\label{sec:conclusion}
		Based on ATLAS-A, an empirical stellar spectral library, we use SLAM to build SSF, a tool that can get arbitrary empirical spectra of stars given arbitrary stellar labels. According to the parameters ranges, we obtain 4  sub-models, SSF-N, -gM, -dM, and -B, with training datasets of different types of stars. 
		
		SSF-N mainly covers the parameter space such that \teff~ from 3700 to 8700 K, \logg~from 0 to 6 dex and [M/H] from - 1.5 to 0.5 dex. SSF-gM covers M giant stars with \teff~ from 3520 to 4000 K and [M/H] from - 1.5 to 0.5 dex. SSF-dM covers M dwarf stars with \teff~from 3295 to 4040 K and {[}M/H{]} from -1.0 to 0.1 dex. SSF-B mainly covers the hot stars with \teff~from 9000 to 24000 K and $M_G$ from -5.2 to 1.5 mag. These spectral libraries can not only provide help for stellar population research, but also can be used to estimate stellar parameters for all types of spectra.
		
		We validate the four libraries by comparing with the known empirical libraries: MILES and MaStar. The photometry can be obtained by integrating the predicted spectra and is compared with the observed photometric information in $g$, $r$ and $i$ bands either from PS1 or from APASS. The predicted spectra well agree with the corresponding spectra from the known libraries and the magnitude calculated from predicted spectra are also well consistent with the corresponding observed ones.

		\begin{acknowledgements}
			This work is supported by the National Key R\&D Program of China No. 2019YFA0405500. C.L. Thanks the National Natural Science Foundation of China (NSFC) with grant Nos.11835057. Guoshoujing Telescope (the Large Sky Area Multi-Object Fiber Spectroscopic Telescope LAMOST) is a National Major Scientific Project built by the Chinese Academy of Sciences, Funding for the project has been provided by the National Development and Reform Commission. LAMOST is operated and managed by the National Astronomical Observatories, the Chinese Academy of Sciences.
		\end{acknowledgements}

		\label{lastpage}
		

\begin{thebibliography}{}
			
			\bibitem[Bagnulo et al.(2003)]{bagn03} Bagnulo, S., Jehin, E., Ledoux, C., et al.\ 2003, The Messenger, 114, 10
			
			\bibitem[Bruzual \& Charlot(2003)]{Bru2003} Bruzual, G., \& Charlot, S.\ 2003, \mnras, 344, 1000
			
			\bibitem[Bu, \& Pan(2015)]{2015MNRAS.447..256B} Bu, Y., \& Pan, J.\ 2015, \mnras, 447, 256
			
			\bibitem[Cardelli et al.(1989)]{car89} Cardelli, J. A., Clayton, G. C., \& Mathis, J. S. 1989, ApJ, 345, 245, (CCM)
			
			\bibitem[Chambers  et al.(2016)]{cha16} Chambers, K. C., Magnier, E. A., Metcalfe, N., et al. 2016, arXiv:1612.05560
			
			\bibitem[\protect\citeauthoryear{Chang \& Lin}{2011}]{Chang2011} Chang., C.-C., Lin., C.-J., 2011, ACM Transactions on Intelligent Systems and Technology, 2:27:1
			
			\bibitem[\protect\citeauthoryear{Deng et al.}{2012}]{2012RAA....12..735D} Deng, L.-C., Newberg, H.~J., Liu, C., et al.\ 2012, Research in Astronomy and Astrophysics, 12, 735
			
			\bibitem[Du et al.(2019)]{Du19} Du, B., Luo, A.-L., Zuo, F., et al.\ 2019, \apjs, 240, 10
			
			\bibitem[Freeman(2012)]{2012ASPC..458..393F} Freeman, K.~C.\ 2012, Galactic Archaeology: Near-Field Cosmology and the Formation of the Milky Way, 458, 393
			
			\bibitem[Green et al.(2019)]{2019ApJ...887...93G} Green, G.~M., Schlafly, E., Zucker, C., et al.\ 2019, \apj, 887, 93. doi:10.3847/1538-4357/ab5362
			
			\bibitem[\protect\citeauthoryear{Gilmore et al.}{2012}]{Gilmore2012}Gilmore G., et al., 2012, Msngr, 147, 25
			
			\bibitem[Guo et al.(2021)]{2021arXiv211006246G} Guo, Y., Zhang, B., Liu, C., et al.\ 2021, arXiv:2110.06246
			
			\bibitem[Henden \& Munari(2014)]{2014CoSka..43..518H} Henden, A. \& Munari, U.\ 2014, Contributions of the Astronomical Observatory Skalnate Pleso, 43, 518
			
			\bibitem[Henden et al.(2016)]{2016yCat.2336....0H} Henden, A.~A., Templeton, M., Terrell, D., et al.\ 2016, VizieR Online Data Catalog, II/336
			
			\bibitem[Ji et al.(2023)]{2023ApJS..265...61J} Ji, W., Liu, C., Deng, L., et al.\ 2023, \apjs, 265, 61. doi:10.3847/1538-4365/acbf42
			
			\bibitem[Koleva et al.(2009)]{kol09} Koleva, M., Prugniel, P., Bouchard, A., et al.\ 2009, \aap, 501, 1269
			
			\bibitem[Li et al.(2014)]{2014ApJ...790..105L} Li, X., Wu, Q.~M.~J., Luo, A., et al.\ 2014, \apj, 790, 105
			
			\bibitem[Li et al.(2021)]{2021ApJS..253...45L} Li, J., Liu, C., Zhang, B., et al.\ 2021, \apjs, 253, 45. doi:10.3847/1538-4365/abe1c1
			
			\bibitem[Liu et al.(2012)]{2012MNRAS.426.2463L} Liu, C., Bailer-Jones, C.~A.~L., Sordo, R., et al.\ 2012, \mnras, 426, 2463. doi:10.1111/j.1365-2966.2012.21797.x
			
			\bibitem[Liu et al.(2014)]{2014ApJ...790..110L} Liu, C., Deng, L.-C., Carlin, J.~L., et al.\ 2014, \apj, 790, 110. doi:10.1088/0004-637X/790/2/110
			
			\bibitem[Le Borgne et al.(2003)]{LEB03} Le Borgne, J.-F., Bruzual, G., Pell{\'o}, R., et al.\ 2003, \aap, 402, 433
			
			\bibitem[Le Borgne et al.(2004)]{2004A&A...425..881L} Le Borgne, D., Rocca-Volmerange, B., Prugniel, P., et al.\ 2004, \aap, 425, 881. doi:10.1051/0004-6361:200400044
			
			\bibitem[Lu, \& Li(2015)]{2015MNRAS.452.1394L} Lu, Y., \& Li, X.\ 2015, \mnras, 452, 1394
			
			\bibitem[Maraston \& Str{\"o}mb{\"a}ck(2011)]{mar11} Maraston, C., \& Str{\"o}mb{\"a}ck, G.\ 2011, \mnras, 418, 2785
			
			\bibitem[\protect\citeauthoryear{Majewski}{2012}]{Majewski2012}Majewski S.~R., 2012, AAS, 219, 205.06 
			
			\bibitem[Ness et al.(2015)]{2015ApJ...808...16N} Ness, M., Hogg, D.~W., Rix, H.-W., et al.\ 2015, \apj, 808, 16. doi:10.1088/0004-637X/808/1/16
			
			\bibitem[Prugniel \& Soubiran(2001)]{PRU01} Prugniel, P., \& Soubiran, C.\ 2001, \aap, 369, 1048
			
			\bibitem[Pickles(1985)]{pickl85} Pickles, A.~J.\ 1985, \apjs, 59, 33
			
			\bibitem[Pickles(1998)]{pickl98} Pickles, A.~J.\ 1998, \pasp, 110, 863
			
			\bibitem[Prugniel et al.(2007)]{2007IAUS..241...68P} Prugniel, P., Koleva, M., Ocvirk, P., et al.\ 2007, Stellar Populations as Building Blocks of Galaxies, 241, 68. doi:10.1017/S1743921307007454
			\bibitem[Qiu et al.(2023)]{2023arXiv230308344Q} Qiu, D., Tian, H., Li, J., et al.\ 2023, arXiv:2303.08344. doi:10.48550/arXiv.2303.08344
			
			\bibitem[S{\'a}nchez-Bl{\'a}zquez et al.(2006)]{sanch06} S{\'a}nchez-Bl{\'a}zquez, P., Peletier, R.~F., Jim{\'e}nez-Vicente, J., et al.\ 2006, \mnras, 371, 703
			
			\bibitem[Sharma et al.(2016)]{2016A&A...585A..64S} Sharma, K., Prugniel, P., \& Singh, H.~P.\ 2016, \aap, 585, A64. doi:10.1051/0004-6361/201526111
			
			\bibitem[\protect\citeauthoryear{Smola \& Sch{\"o}lkopf}{2004}]{ss2004} Smola, A. J., Sch{\"o}lkopf, B, Statistics and Computing, 14, 199
			
			\bibitem[Vazdekis et al.(2010)]{vaz10} Vazdekis, A., S{\'a}nchez-Bl{\'a}zquez, P., Falc{\'o}n-Barroso, J., et al.\ 2010, \mnras, 404, 1639
			
			\bibitem[Vazdekis et al.(2012)]{vaz12} Vazdekis, A., Ricciardelli, E., Cenarro, A.~J., et al.\ 2012, \mnras, 424, 157
			
			\bibitem[Valdes et al.(2004)]{vald04} Valdes, F., Gupta, R., Rose, J.~A., et al.\ 2004, \apjs, 152, 251
			
			\bibitem[Yan et al.(2019)]{yan19} Yan, R., Chen, Y., Lazarz, D., et al.\ 2019, \apj, 883, 175
			
			\bibitem[Zhang et al.(2020)]{2020ApJS..246....9Z} Zhang, B., Liu, C., \& Deng, L.-C.\ 2020, \apjs, 246, 9. doi:10.3847/1538-4365/ab55ef
			
			
			
			
			
			
		\end{thebibliography}
	\end{document}